\documentstyle[epsf]{mn}

\newif\ifAMStwofonts

\newcommand{\ra}[4]{$#1^{\mathrm{h}}#2^{\mathrm{m}}#3^{\mathrm{s}}\!\!.#4$}
\newcommand{\dec}[3]{$#1^{\circ}#2^{\prime}#3^{\prime\prime}\!\!$}

\ifoldfss
  \ifCUPmtlplainloaded \else
    \NewTextAlphabet{textbfit} {cmbxti10} {}
    \NewTextAlphabet{textbfss} {cmssbx10} {}
    \NewMathAlphabet{mathbfit} {cmbxti10} {} 
    \NewMathAlphabet{mathbfss} {cmssbx10} {} 
  \fi
  \ifAMStwofonts
    \ifCUPmtlplainloaded \else
      \NewSymbolFont{upmath} {eurm10}
      \NewSymbolFont{AMSa} {msam10}
      \NewMathSymbol{\upi}     {0}{upmath}{19}
      \NewMathSymbol{\umu}     {0}{upmath}{16}
      \NewMathSymbol{\upartial}{0}{upmath}{40}
      \NewMathSymbol{\leqslant}{3}{AMSa}{36}
      \NewMathSymbol{\geqslant}{3}{AMSa}{3E}

    \fi
  \fi
\fi 

\ifnfssone
  \newmathalphabet{\mathit}
  \addtoversion{normal}{\mathit}{cmr}{m}{it}
  \addtoversion{bold}{\mathit}{cmr}{bx}{it}
  \newmathalphabet{\mathbfit} 
  \addtoversion{normal}{\mathbfit}{cmr}{bx}{it}
  \addtoversion{bold}{\mathbfit}{cmr}{bx}{it}
  \newmathalphabet{\mathbfss} 
  \addtoversion{normal}{\mathbfss}{cmss}{bx}{n}
  \addtoversion{bold}{\mathbfss}{cmss}{bx}{n}
  \ifAMStwofonts
    \ifCUPmtlplainloaded \else
      \UseAMStwoboldmath
      \makeatletter
      \new@mathgroup\upmath@group
      \define@mathgroup\mv@normal\upmath@group{eur}{m}{n}
      \define@mathgroup\mv@bold\upmath@group{eur}{b}{n}
      \edef\UPM{\hexnumber\upmath@group}
      \new@mathgroup\amsa@group
      \define@mathgroup\mv@normal\amsa@group{msa}{m}{n}
      \define@mathgroup\mv@bold\amsa@group{msa}{m}{n}
      \edef\AMSa{\hexnumber\amsa@group}
      \makeatother
      \mathchardef\upi="0\UPM19
      \mathchardef\umu="0\UPM16
      \mathchardef\upartial="0\UPM40
      \mathchardef\leqslant="3\AMSa36
      \mathchardef\geqslant="3\AMSa3E
    \fi
  \fi
\fi 

\ifnfsstwo
  \DeclareMathAlphabet{\mathbfit}{OT1}{cmr}{bx}{it}
  \SetMathAlphabet\mathbfit{bold}{OT1}{cmr}{bx}{it}
  \DeclareMathAlphabet{\mathbfss}{OT1}{cmss}{bx}{n}
  \SetMathAlphabet\mathbfss{bold}{OT1}{cmss}{bx}{n}
  \ifAMStwofonts
    \ifCUPmtlplainloaded \else
      \DeclareSymbolFont{UPM}{U}{eur}{m}{n}
      \SetSymbolFont{UPM}{bold}{U}{eur}{b}{n}
      \DeclareSymbolFont{AMSa}{U}{msa}{m}{n}
      \DeclareMathSymbol{\upi}{0}{UPM}{"19}
      \DeclareMathSymbol{\umu}{0}{UPM}{"16}
      \DeclareMathSymbol{\upartial}{0}{UPM}{"40}
      \DeclareMathSymbol{\leqslant}{3}{AMSa}{"36}
      \DeclareMathSymbol{\geqslant}{3}{AMSa}{"3E}
    \fi
  \fi
\fi 

\ifCUPmtlplainloaded \else
  \ifAMStwofonts \else 
    \def\upi{\pi}
    \def\umu{\mu}
    \def\upartial{\partial}
  \fi
\fi

\title{The Infrared Jet in 3C\,66B}
\author[D. Tansley, M. Birkinshaw, M. J. Hardcastle and D. M. Worrall]
       {D. Tansley, M. Birkinshaw, M. J. Hardcastle and D. M. Worrall\\
        Department of Physics, University of Bristol, Tyndall Avenue, Bristol, BS8 1TL}
\date{ }

\pagerange{\pageref{firstpage}--\pageref{lastpage}}
\pubyear{1999}

\begin{document}

\maketitle

\label{firstpage}

\begin{abstract}
	
\noindent We present images of infrared emission from the radio jet in
3C\,66B. Data at three wavelengths (4.5, 6.75 and 14.5 $\mu \rm{m}$)
were obtained using the Infrared Space Observatory. The
6.75 $\mu \rm{m}$ image clearly shows an extension aligned with the
radio structure.  The jet was also detected in the 14.5 $\mu \rm{m}$
image, but not at 4.5 $\mu \rm{m}$. The radio-infrared-optical
spectrum of the jet can be interpreted as synchrotron emission from a
population of electrons with a high-energy break of 4$\times$10$^{11}$
eV.  We place upper limits on the IR flux from the radio counter-jet.
A symmetrical, relativistically beamed twin-jet structure is
consistent with our results if the jets consist of multiple
components.

\end{abstract}

\begin{keywords}
galaxies: active; galaxies: jets; galaxies: individual: 3C\,66B; 
infrared: galaxies  
\end{keywords}

\section{Introduction}

Jets are common in extragalactic radio sources.  They are believed to
trace beams of particles and fields ejected from the compact central
region at bulk velocities approaching the speed of light, combined
with material swept up by the flow, and are usually visualised through
the loss of energy as synchrotron radiation.  The emission measured at
radio frequencies should, in the absence of a high-energy cutoff in
the electron energy spectrum, extend to the optical and infrared
bands, although optical counterparts are detected for relatively few jets 
[Martel et al. (1998) list 9 such cases] and an even smaller number of infrared
jets have been reported. One noteable exception is the jet in
M87 which has been detected in the near-IR (Killeen et al. 1984) 
and with ISO in the mid-IR (Tansley et al. in preparation).

The radio galaxy 3C\,66B has one of the best-studied optical jets. The
source has a redshift of 0.0212 (Stull et al. 1975) and is associated
with a 13th magnitude elliptical galaxy in a small group in the
vicinity of the cluster Abell 347. Its radio structure (Leahy, J\"agers and Pooley 1986; 
Hardcastle et al. 1996) has been studied in detail, and reveals a morphology which
is intermediate between an edge-darkened double and a head-tail
structure, possibly indicating the effect of the parent galaxy's
motion through the ambient medium.
 
The optical jet in 3C\,66B was first detected by Butcher, van Breugel
and Miley (1980).  Further ground-based work (Fraix-Burnet, Nieto and
Poulain 1989) has revealed the presence of five distinct knots of
polarized optical emission along the jet.  These correspond to five
similar knots in the radio emission. A tentative optical detection of
the southern counter-jet was made in the I-band (Fraix-Burnet
1997). This is interesting as it allows the radio-optical spectra in
the jet and the counter-jet to be compared. If the simplest
twin-relativistic-beam model is valid, then the spectra of the jet and
counter-jet should differ in intensity and shape in ways described by
a single component synchrotron model.

Hubble Space Telescope (\textit{HST}) images of 3C\,66B (Macchetto et al. 1991) show the optical jet on
scales of $\sim 0.1$ arcsec, and have revealed an intriguing
double-stranded filamentary structure which resembles the
radio image at the same resolution. A comparison between \textit{HST} and
VLA data at 0.25 arcsec resolution (Jackson et al. 1993) shows further
similarities.

Here we present data obtained with the Infrared Space
Observatory\footnote{ISO is an ESA project with instruments funded by
ESA Member States (especially the PI countries: France, Germany, the
Netherlands and the United Kingdom) and with the participation of ISAS
and NASA.}  (\textit{ISO}) which appear to detect the jet at 6.75 and
14.5 $\mu \rm{m}$.

Throughout this paper we assume H$_{0}$=50 km s$^{-1}$ Mpc$^{-1}$.

\section{Observations and Data Reduction}

3C\,66B was observed with the European Space Agency telescope
\textit{ISO} as part of a larger project to study infrared emission
from the dusty central regions of a sample of 3CRR objects (Tansley et
al. in preparation).  Observations were carried out with the ISOCAM
instrument (\textit{ISO's} near-mid IR imaging camera; Kessler et
al. 1996, Cesarsky et al. 1996) using the LW1, LW2 and LW3 filters to
obtain images in bands 4.00-5.00, 5.00-8.50 and 12.00-18.00 $\mu$m
centred on wavelengths of 4.5, 6.75 and 14.5 $\mu \rm{m}$,
respectively.  A pixel scale of 3 arcsecs was selected, and the
angular resolution of the data was 3.26, 3.98 and 4.03 arcsecs for the three filters 
(measured from the supplied PSF libraries),
LW1, LW2 and LW3 respectively. The noise
levels on the final images are close to the theoretically expected
values of 69, 71, 70 $\mu$Jy for the three filters, respectively.

The data were processed initially using the ISOCAM Consortium/ESA CAM
Interactive Analysis (CIA) package (Ott et al. 1997), running under
the Interactive Data Language, IDL. The data were corrected for
instrumental and cosmic-ray glitch artifacts and also for data
stabilization using the multi-resolution median method.\footnote{For
details see: http://isowww.estec.esa.nl/manuals/\\HANDBOOK/III/cam\_hb/node51.html.}
The most recent (15th March, 1998) calibration-library flat-field and dark frames were then applied.  The
final calibrated images have unexpectedly ``lumpy'' backgrounds,
perhaps indicating that the supplied flat-fields were not ideal. Further
details of the analysis are given in Tansley et al. (in preparation)

\begin{table*}
\begin{center}
\begin{tabular}{r | r |r |r |} \hline

Components 		& Component Flux Density	& Goodness of Fit				\\
			& mJy				& $\chi ^{2}$ / degrees of freedom	\\	\hline
			&				&					\\
AGN		   	&4.4$\pm$0.5				& 450.1/60					\\
			&				&					\\
Galaxy			&8.4$\pm$0.4			& 208.8/58				\\
			&				&					\\
AGN  		 	&2.0$\pm$0.3			& 157.7/55				\\
+ Galaxy		&8.2$\pm$0.4			&					\\
			&				&					\\
One jet point  		&0.4$\pm$0.1			& 143.3/52				\\
+ AGN		   	&1.8$\pm$0.3			&					\\
+ Galaxy		&8.3$\pm$0.4			&					\\
			&				& 					\\
Two jet points 		&0.8$\pm$0.1			& 135.2/49				\\
+ AGN		   	&0.3$\pm$0.1			& 					\\
+ Galaxy		&2.1$\pm$0.2			&					\\ 
			&8.2$\pm$0.4			&					\\ \hline
\end{tabular}
\caption{Goodness of fit comparison for variations of the model fit at 6.75 $\mu$m.} \label{chis}
\end{center}
\end{table*}

\begin{table*}
\begin{center}
\begin{tabular}{r | r | r|} \hline
Component	&	Position of Component		& Offset From Centre	\\
		&	RA \& Dec			& of Galaxy/ arcsecs	\\ \hline
		&					&			\\
Galaxy		&\ra{02}{23}{11}{2}, \dec{42}{59}{28}	& N/A			\\
		&					&			\\
AGN		&\ra{02}{23}{11}{2}, \dec{42}{59}{28}	& 0			\\
		&					&			\\
Jet 1		&\ra{02}{23}{12}{0}, \dec{42}{59}{35}	& 13.9 			\\
		&					&			\\
Jet 2		&\ra{02}{23}{11}{6}, \dec{42}{59}{32}	& 7.2			\\ \hline

\end{tabular}
\caption{Positions and offsets of best-fit components.} \label{pos}
\end{center}
\end{table*}

\section{Results and Analysis}

\begin{figure*}
  \begin{center}
 \leavevmode
 \epsfxsize 8cm
	\epsfbox{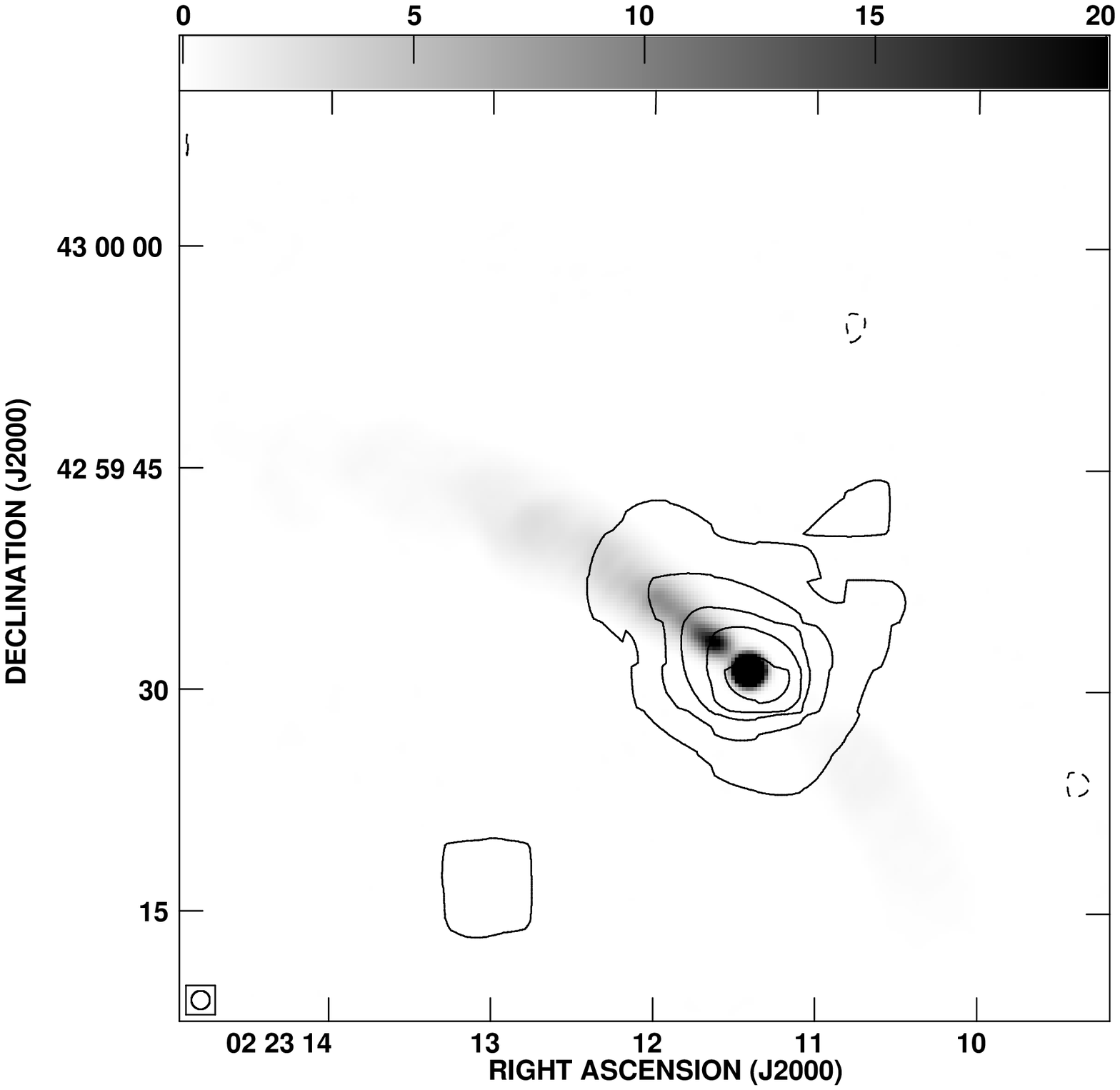}
 \epsfxsize 8cm
	\epsfbox{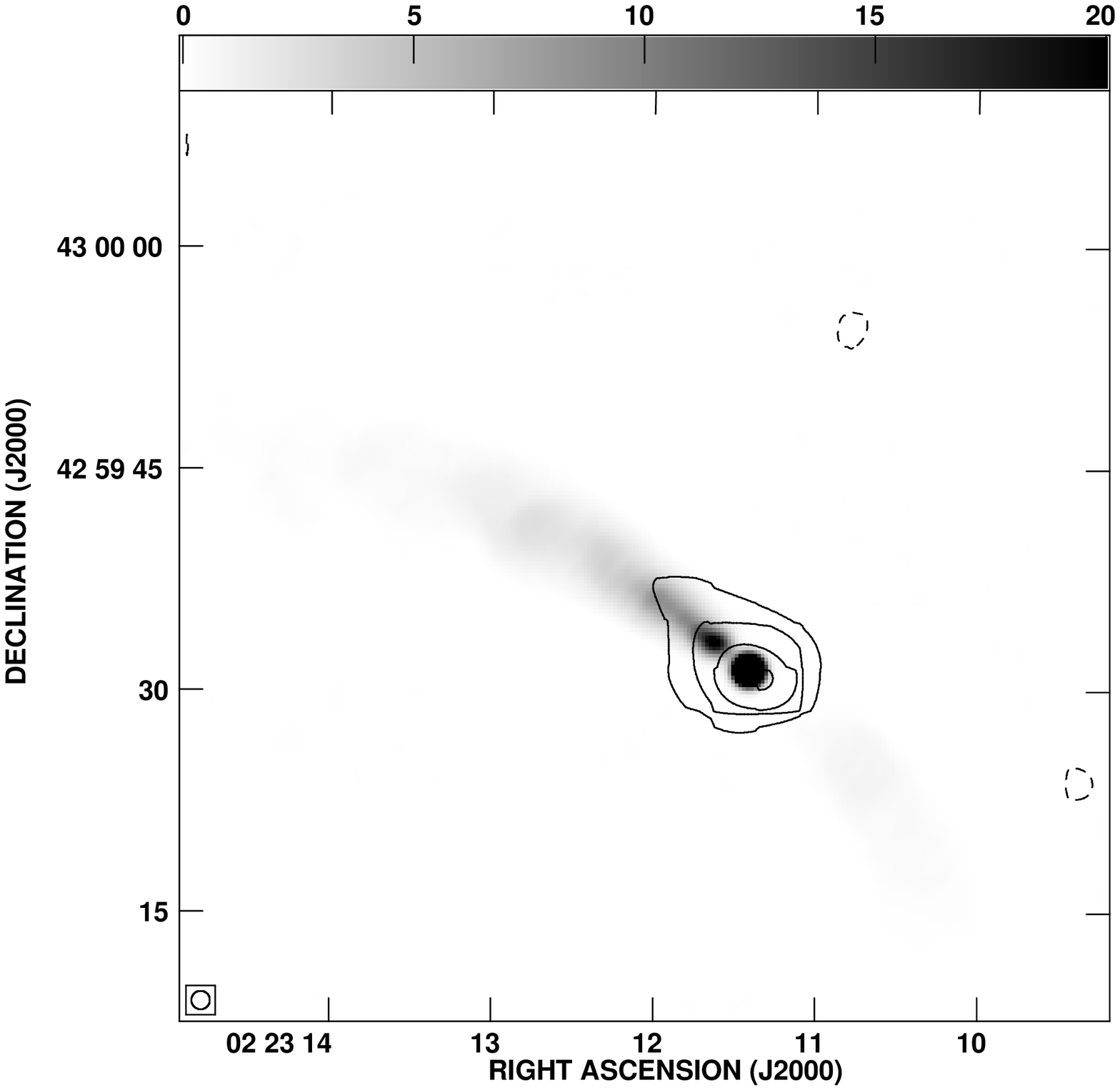}
  \end{center}
 \caption{Image showing the \textit{ISO} LW2 data, with galaxy emission included (left) and subtracted (right), as contours (lowest contour at 
3$\sigma$; levels 0.11mJy/pixel $\times$ -1, 1, 2, 3, 4, 5), with VLA image at 1425 MHz (Hardcastle et al. 1996) shown in grey-scale (0 to 20 mJy/beam). Note the disappearence of the companion galaxy from the subtracted image (right)} \label{vla}
\end{figure*}

The 6.75 $\mu$m image can be seen in Figure \ref{vla}. The contours of
this image represent the ISO data and the greyscale is a 1425 MHz VLA
image from Hardcastle et al. (1996)

The final calibrated images were fitted with a three-component model,
consisting of an extended galactic component (modelled as an
elliptical Gaussian based on the optical galaxy as a template), an
unresolved AGN component, and a set of one or two unresolved points representing
emission from the jet component. The goodness of fit of the model at
various stages of complexity is shown for the LW2 data in Table
\ref{chis}. Table \ref{pos} shows the positions and central offsets of the 
best fit components with this filter.
Fitting only an extended galactic component or unresolved
AGN component gives a significantly worse fit (by $\Delta\chi^{2}$=51
and 292 respectively) than a combination of the two. At this stage, a residual map
of the \textit{ISO} LW2 (and LW3) data showed clear evidence for
excess emission, with structure significantly different from, and
brighter than, the background ``lumps''. This excess was to the east of
the AGN/galaxy position in the region of the inner radio and optical
jet. The overall fit improves further (by $\Delta\chi^{2}$=14.4 or 22.5)
when one or two unresolved components representing jet emission were added to the model.

Table \ref{results} presents the fitted flux densities in the three
\textit{ISO} filters for the jet components based on the positions derived
from the LW2 image. We
believe the high $\chi^{2}$ values to be caused primarily by
inaccuracies in the flat-field calibration frame rather than the model
itself: a $\chi^{2}$ map of the fitted region is similar to a
$\chi^{2}$ map of an off-source region in showing patchy (scale
$\sim$6 - 9 arcsecs) low-level (peaks of $\sim$0.1 mJy) residual
structure. Although this structure lies above the nominal noise level, it 
is much fainter than the jet components that we identify: if a post-facto
assessment of the image noises is made, based on the final $\chi^{2}$ values, the
errors on the LW1 and LW2 flux densities would increase by 50 and 68\% respectively.
Errors in the absolute flux scale can also contribute to uncertainties in the 
measured flux densities. The \textit{ISOCAM} documentation quotes an uncertainty
in the spectral energy distribution calibration of $\pm$5\%. Which is much less than
the random errors on the jet and is not accounted for in our quoted errors.

Table \ref{fluxes} lists the jet flux densities at radio, IR and optical
wavelengths.  The radio measurements were taken from images obtained
with the VLA at 1.4 and 8 GHz (Hardcastle et al. 1996) integrated over
the same regions covered by the IR components.  The corresponding I
band flux density was taken from the galaxy-subtracted image of
Fraix-Burnet (1997) obtained at the Canada-France-Hawaii
telescope. The optical flux densities were taken from the \textit{HST}
measurements of Jackson et al. (1993), or measured directly from
archival \textit{HST} data. However, since the field of view (FOV) of the \textit{HST}
Faint Object Camera (FOC) does not include the full extent of the IR
emission, we assumed that the radio-optical spectral index of the
entire region was equal to that of the observed region and we scaled the
measured optical flux densities by a factor 1.66 to correct for the
40\% of the radio emission that lies outside the FOC's FOV.

\begin{table*}
\begin{center}
\begin{tabular}{l | c r r r |} \hline

Filter	&	Goodness of Fit			& \multicolumn{3}{c |} {Flux Densities}               \\
	& $\chi ^{2}$ / degrees of freedom 	& Component 1	& Component 2	& Total  	      \\	
	&					& (mJy)		& (mJy)		& (mJy)		      \\ \hline
LW1$\dagger$&		110.0/55		& $<$0.36	& $<$0.36	& $<$0.72	      \\
	&					&		&		&		      \\
LW2	&		135.2/49		& 0.43$\pm$0.13	& 0.67$\pm$0.18	& 1.10$\pm$0.22	      \\
	&					&		&		&		      \\
LW3	&		116.0/49		& 1.08$\pm$0.26	& 0.59$\pm$0.18	&1.66$\pm$0.31	      \\
	&					&		&		&		      \\ \hline

\end{tabular}
\caption{Flux densities for the infrared jet emission. $\dagger$ Jet not detected with this
filter. 3$\sigma$ upper limits are based on a model without jet components. Model fitted without jet components.} \label{results}
\end{center}
\end{table*}

\begin{table*}
\begin{center}
\begin{tabular}{l | c c |} \hline

Frequency		&	Flux Density			&	Reference		\\
Hz			&	mJy				&				\\ \hline
			&					&				\\
1.30$\times$10$^{15}$	&	0.0061	$\pm$  0.0009		& Jackson et al. 1993 (scaled to\\
			&					& account for smaller FOV)	\\
9.67$\times$10$^{14}$	&	0.0102	$\pm$  0.0016		& Jackson et al. 1993 (scaled to\\
			&					& account for smaller FOV)	\\
8.80$\times$10$^{14}$	&	0.0128	$\pm$  0.0014		& Macchetto et al. 1991		\\
			&					&				\\
7.32$\times$10$^{14}$	&	0.0184	$\pm$  0.0024		& This paper (from public archive)\\
			&					&				\\
3.33$\times$10$^{14}$	&	0.111	$\pm$  0.001		& Fraix-Burnet 1997		\\
			&					&				\\
6.67$\times$10$^{13}$	&	$<$0.72				& This paper			\\
			&					&				\\
4.44$\times$10$^{13}$	&	1.10	$\pm$  0.22		& This paper			\\
			&					&				\\
2.06$\times$10$^{13}$	&	1.66	$\pm$  0.31		& This paper			\\
			&					&				\\ 
8.41$\times$10$^{9}$	&	105.0	$\pm$  2.7		& Hardcastle et al. 1996	\\
			&					&				\\
1.42$\times$10$^{9}$	&	315.0	$\pm$  9.0	 	& Hardcastle et al. 1996	\\
			&					&				\\\hline

\end{tabular}
\caption{Optical, infrared and radio fluxes for the inner jet.} \label{fluxes}
\end{center}
\end{table*}

\begin{figure*}
  \begin{center}
 \leavevmode
 \epsfxsize 8cm
	\epsfbox{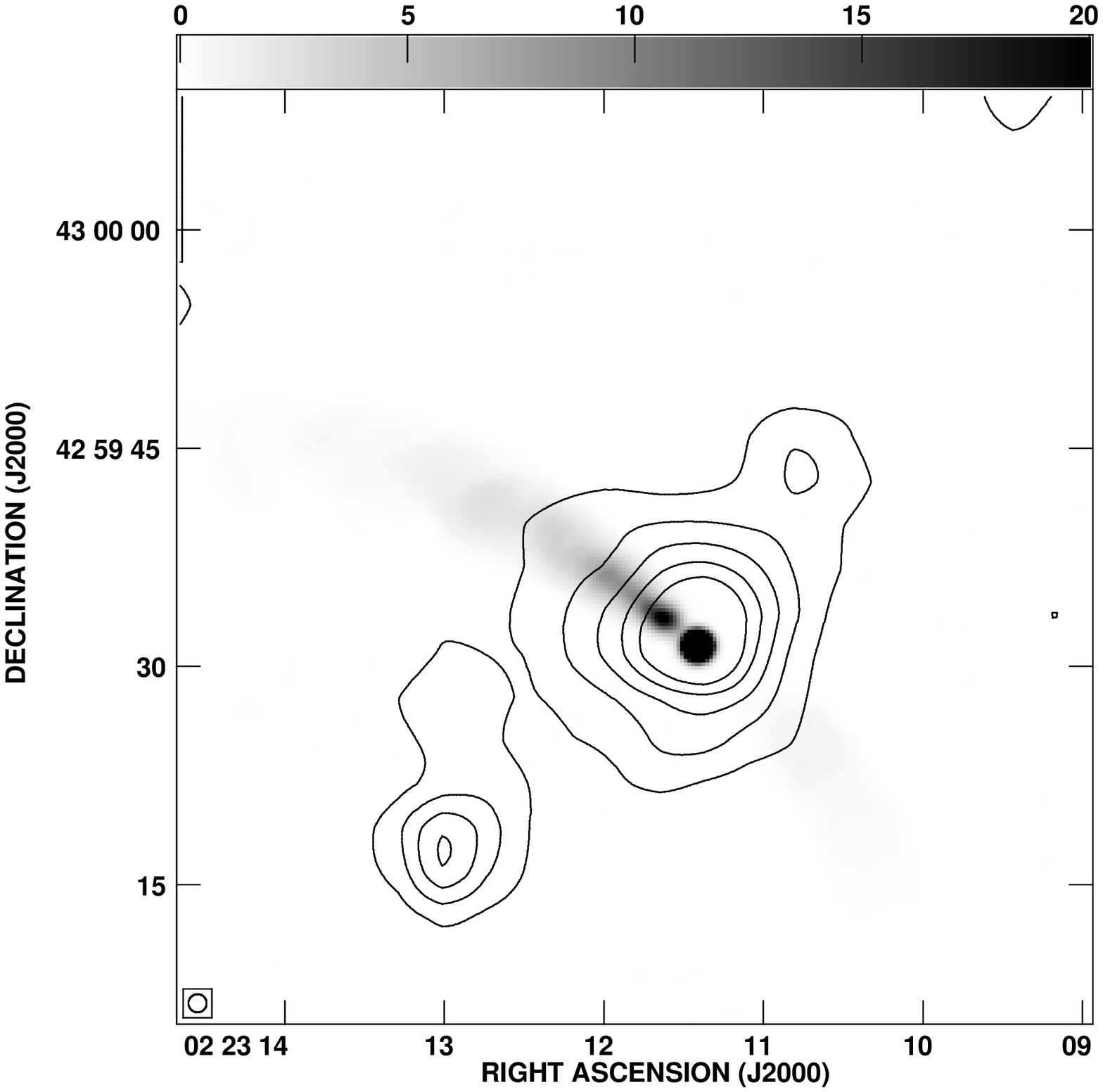}
 \epsfxsize 8cm
	\epsfbox{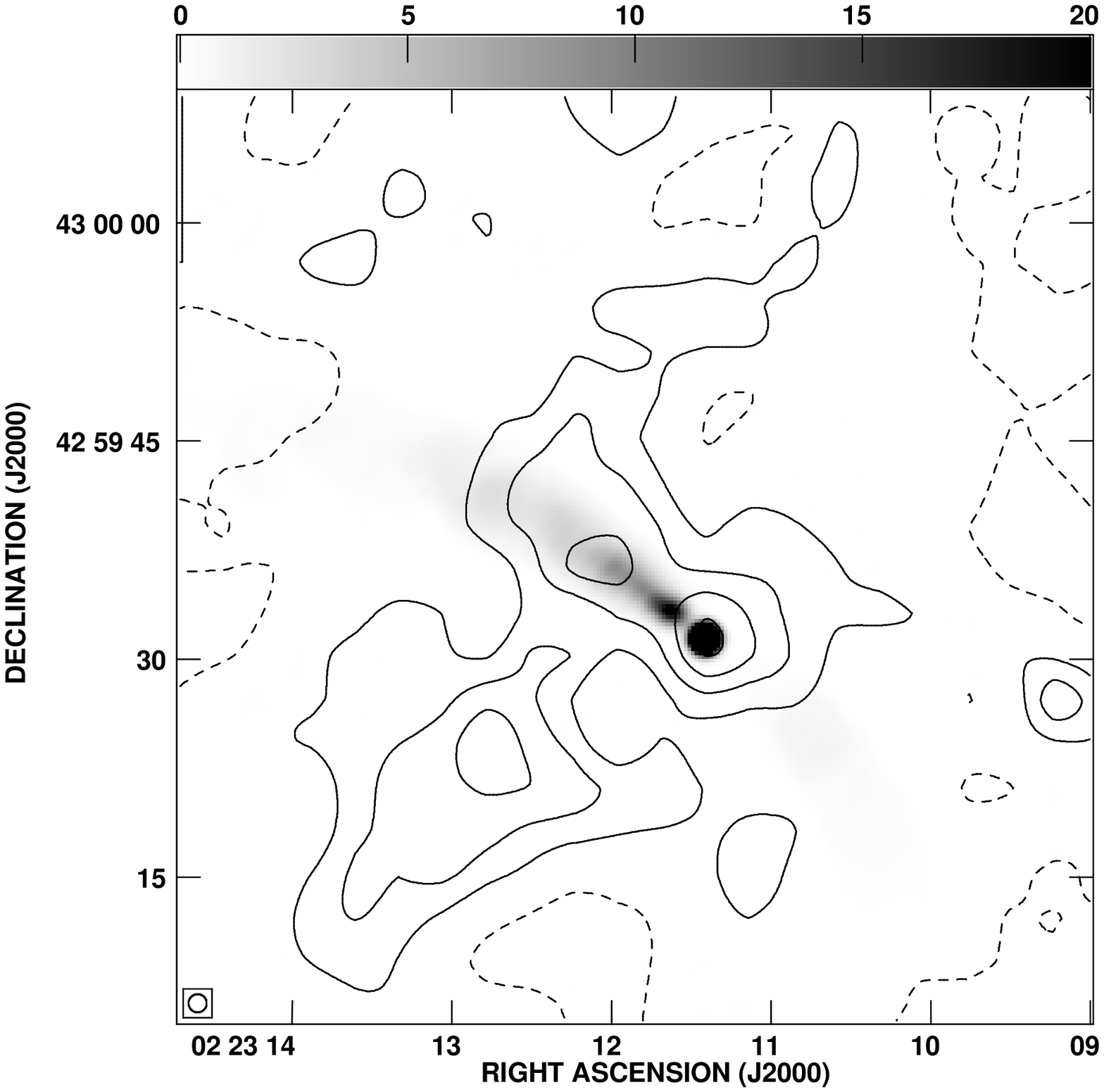}
  \end{center}
 \caption{Image showing the \textit{ISO} LW1 (left) and LW3 (right) data as contours (lowest contour at 
3$\sigma$ [0.09mJy/pixel and 0.18mJy/pixel $\times$ -1, 1, 2, 3, 4, 5 respectively] level), with VLA image at 1425 MHz 
(Hardcastle et al. 1996) shown in grey-scale (0 to 20 mJy/beam). The companion galaxy can be seen clearly at \ra{02}{23}{13}{0}, \dec{42}{59}{17}.} \label{vla2}
\end{figure*}

\begin{figure*}
  \begin{center}
 \leavevmode
 \epsfxsize 10cm
	\epsfbox{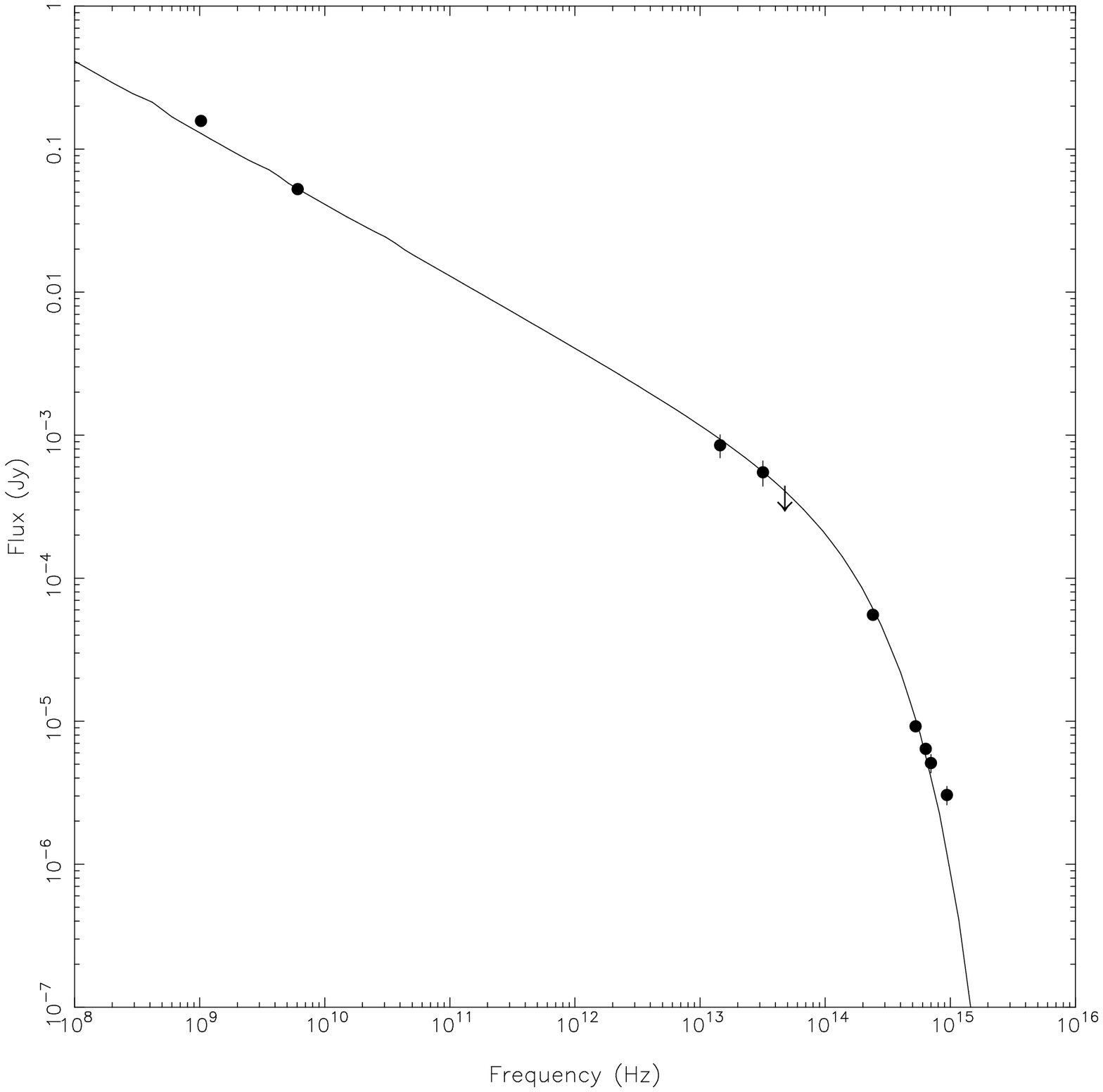}
 \epsfxsize 10cm
 	\epsfbox{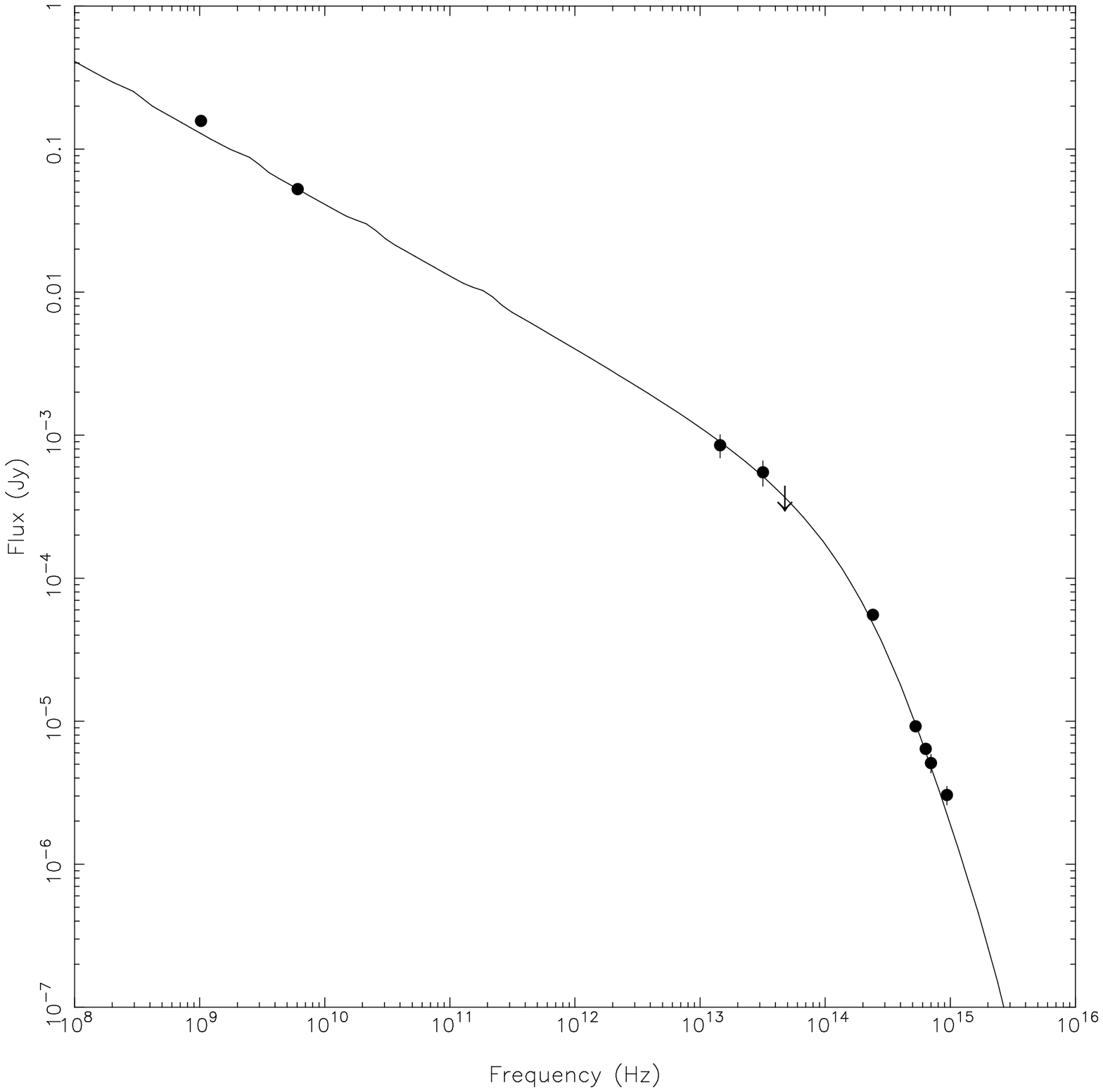}
  \end{center}
 \caption{Optical-IR-Radio spectrum of 3C\,66B showing fitted synchrotron profile. Top: calculated using an electron spectrum with a 
high-energy cutoff. Bottom: calculated using an electron spectrum with a high-energy break. In both cases the frequencies and the flux densities are in the 
rest frame of the inner-jet component, taking $\beta$=0.7 and $\theta$=45$^{\circ}$ (see Section 4).} \label{spectrum}
\end{figure*}

\section{Discussion}

Figure \ref{spectrum} shows the measurements from Table \ref{fluxes}
fitted with two separate synchrotron models. In both cases, the volume 
was assumed to be a simple cylinder (of length 5 kpc and radius 0.5 kpc) 
along the length of the IR jet. We assumed that the emission arises from 
a plasma with velocity v=0.7c oriented at 45$^\circ$ to the line of sight. These assumptions
have only a modest effect on the best-fit field, and are consistent with the 
beam decomposition discussed later (Section 4).
 
The simplest spectral shape that fits the data is one in which p=2 (where $N(E) dE =
N_{0} E^{-p}$) in
the radio-optical region that turns down sharply at higher energies to match
the optical points. We fitted two synchrotron spectra based on indicative models 
of the population of relativistic electrons. The first model population has a high energy cutoff at
E$_{max}$=6.5$\times$10$^{11}$eV and an equipartition magnetic field of 3.7 nT. This fit reproduces
the optical points badly, giving a spectrum that is too steep. The second model
consists of a broken power law population of electrons, where the spectrum breaks from p=2 to p=6 at an energy 
E$_{break}$=4.5$\times$10$^{11}$eV. The maximum energy in this model is
E$_{max}$=15$\times$10$^{11}$eV, and the same equipartition field of 3.7 nT is present. 
The broken-power-law model shows a considerably better fit to the optical points
($\chi^{2}$=215 as compared to $\chi^{2}=391$ for the model with no power-law break).

The worst fitting point in both models is the low-frequency radio point, though
we attribute this to the presence of steep-spectrum material in the jet, possibly indicative 
of a two-component flow in the jet structure (discussed later). Similarly, both fits poorly reproduce
the I-band data point, though this may merely be a consequence of deficiencies in the simple models
since the spectrum near I band is strongly dependent on the shape of the break in the electron spectrum. 
Of greater interest is the higher-than-expected
UV point from the \textit{HST} observation as this may represent an unmodelled component of the electron
energy distribution.

To test whether the flatter optical spectrum is an artifact of our
extrapolation of the observed FOC flux densities, fluxes were measured
on the radio, I-band and \textit{HST} images only within the FOV of the
FOC. These data points still show a shallower optical spectrum than is
predicted by our two models. Thus, the extrapolation of the optical flux
over the full radio-jet region is not the cause of the
flattening. Furthermore, a U-band observation of
3C\,66B using the STIS instrument on \textit{HST} (Sparks et al. in preparation), 
covers the full extent
of the radio jet and has a flux density that is similar to the scaled
Jackson et al. (1993) flux density in the same band.

Forthcoming Chandra X-ray observations of this source will allow us to
constrain further the model parameters of the emission; in particular
they will show whether the flattening of the spectrum in the optical
continues to higher frequencies. If this is true, it may be the case
that we are seeing emission in the blue dominated by a newer, flat spectrum
component, the origin of which is uncertain. Further observations in the
optical and UV would be needed to investigate this.

The counter-jet seen in radio maps of Hardcastle et al. and possibly
in the optical images (Fraix-Burnet 1997) is not detected with
\textit{ISO}. Upper limits for the flux densities 
(Table \ref{counter}) have been measured
using the radio and I-band maps as guides to the emission region. The
radio-IR spectral index is found to be $\alpha >$ 0.5.

\begin{figure*}
  \begin{center}
 \leavevmode
 \epsfxsize 10cm
	\epsfbox{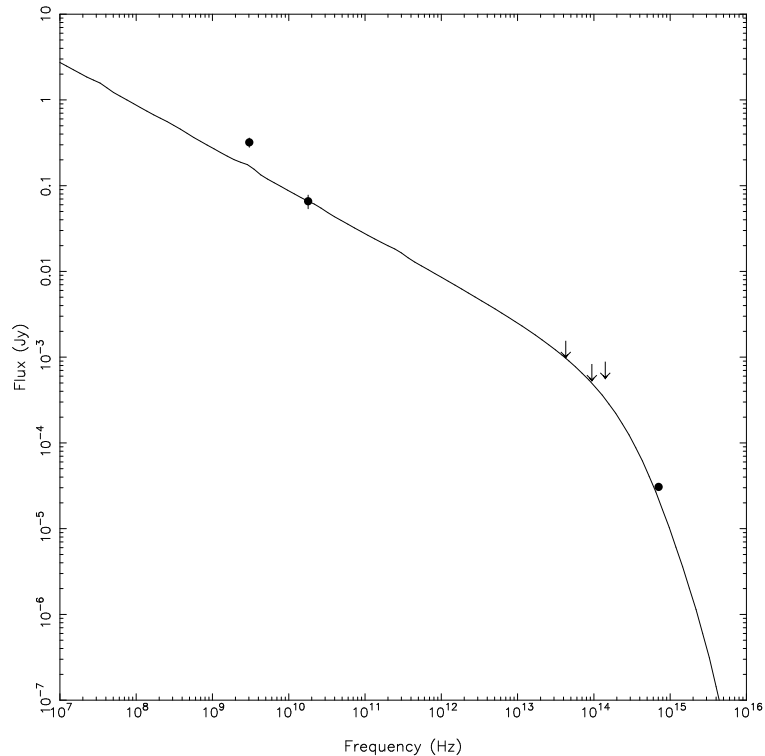}
  \end{center}
 \caption{Radio-IR-Optical spectrum of the counter-jet in 3C\,66B showing fitted high-energy cutoff synchrotron profile. Rest-frame frequencies shown. The frequencies and flux densities shown are in the rest frame of the inner-jet component, taking $\beta$=0.7 and $\theta$=45$^{\circ}$ (see Section 4).} \label{spectrum2}
\end{figure*}

Fraix-Burnet (1997) identified several components of the counter-jet
which appeared coincident with radio features. The measured
radio and optical flux densities of these knots imply that the 
counter-jet has a flatter radio-to-optical spectrum than
the jet. This seems to contradict the usual interpretation of the jet and
counter-jet as physically identical, with their brightness
asymmetry caused by the relativistic Doppler effect. Our IR upper
limits lie above the fluxes that we expect for the counter-jet,
based on a simple broken-power-law model with the
same parameters as the main jet. 
This model is plotted with the data in Figure \ref{spectrum2}.
As can be seen, the fit to the optical point in the counter-jet
spectrum is poor, and agrees with the finding of Fraix-Burnet (1997)
that the spectrum does not conform to that expected if 3C66B was a 
twin beamed structure with identical jets. Instead, it is possible that
either the source is not described by a symmetrical beamed-jet model or
that the observed optical structures are not real.

The radio spectrum of the counter-jet spectrum is 
steeper ($\alpha$ = 0.88) than the radio points of the jet
($\alpha$ = 0.62). A possible explanation could
be that we are seeing the superposition of an old population of
material (forming a sheath around the jet) combined with
newer material recently ejected and passing through the
sheath. Evidence for such an enclosing structure has been reported in
(for example) Hardcastle et al. (1996), Laing (1996) and Hardcastle et al. (1997)
and is also described by O'Dea and Owen (1987) in relation to the radio
maps of NGC 1265. As a weak sheath of emission appears at only 1415 MHz
in NGC 1265, O'Dea and Owen concluded that the sheath has a
steeper spectrum than the jet itself, as would be expected from older
material which had undergone spectral ageing. The presence of the
sheath is attributed to the diffusion of particles out of the beam
into the static external medium.

In this model, we expect the inner component to be travelling faster
than the sheath material, so that emission from the inner flow is
relativistically-boosted on the jet side and dimmed on the counter-jet
side, relative to sheath emission.  Hence the spectrum of the
counter-jet would be less dominated by the inner material,
and consequently steeper. Hardcastle et al. (1996) have
produced a polarization intensity map that shows clearly a distinct
line of depolarization marking the boundary between the sheath and the
inner jet in 3C\,66B. Flux-density measurements of the inner and outer
regions of the jet show that the inner jet is dimmed on the counter-jet side
by a greater amount than the sheath component, as the model predicts.

If we accept this model of the inner beam and outer sheath
then we can use the measured flux densities from the radio
polarization intensity map to estimate the inner flow and sheath
velocities. We can represent the emission from the jet as the sum of
emission components.

\[
S_{total} = \epsilon_{sheath} V_{sheath} \delta_{sheath}^{\alpha_{sheath}+2} + \epsilon_{inner} V_{inner} \delta_{inner}^{\alpha_{inner}+2}
\]

\noindent where $\delta$ is the bulk relativistic Doppler factor:

\[
\delta = \gamma^{-1} (1 - \beta cos\theta)^{-1}
\]

\noindent $\epsilon$ is the emissivity of the material, V is the
volume of the emitting region and $\alpha$ is the spectral index of
the region (either sheath or inner). $\theta$ is the orientation angle
to the line of sight and $\beta$ is the ratio of the velocity to the
speed of light, $c$. Measuring equal volumes on both sides and
equating each component, we find $\beta_{inner} \sim 0.7$ and
$\beta_{sheath} \sim 0.2$ (using $\alpha_{inner}$=0.62 and
$\alpha_{sheath}=0.88$ measured from the jet and counter-jet radio
spectral indices respectively and an orientation angle to the line of sight,
$\theta$ = 45$^{\circ}$).

The observed steepness of the spectrum of the counter-jet in the radio
initially places doubt on the association of the I-band detections
with material in the jet itself. However, if the multi-component model
proposed here is accepted, then at higher frequencies the counter-jet 
may be dominated by the inner jet rather than the
steeper spectrum sheath, despite the relativistic dimming of the inner jet.
Using this assumption, we can perform a simple calculation to test whether
the optical flux of the counter-jet from the Fraix-Burnet image is what we
expect to observe. In fact, using the Doppler parameters of the inner jet, the 
optical spectral index of $\alpha$=1.4 (Jackson et al. 1993) and the 
flux density measurement from the corresponding frequency in the jet, we find an
expected flux density of 2 $\mu$Jy, much lower than the flux of 7 $\mu$Jy measured by
Fraix-Burnet (1997).

\section{Conclusions}

We have made the first mid-IR detection of emission from a radio
jet. Fits to synchrotron models using radio and optical data points
have shown that the overall spectral-energy distribution of 3C\,66B
resembles a simple synchrotron spectrum from the radio to the optical,
though the optical slope is unexpectedly flat, perhaps because of
inhomogeneities in the jet. The counter-jet in 3C\,66B is not detected
in our data. The jet and counter-jet spectra can be interpreted in
terms of a model in which the beam structure is composed of a
slow-moving ($\beta$=0.2), steep-spectrum component and a faster-moving 
($\beta$=0.7), flatter-spectrum component, a conclusion
supported by radio images of polarised emission from the jet structure.

\begin{table}
\begin{center}
\begin{tabular}{l | c c c|} \hline

Frequency		&	Flux Density	&	References	\\
Hz			&	mJy		&			\\ \hline
			&			&			\\
3.33$\times$10$^{14}$	&	0.007		& Measured from the maps of 	\\ 
			&			& Fraix-Burnet 1997		\\
6.67$\times$10$^{13}$	&	$<$0.16		& This paper.		\\
			&			&			\\
4.44$\times$10$^{13}$	&	$<$0.15		& This paper.		\\
			&			&			\\
2.06$\times$10$^{13}$	&	$<$0.29		& This paper.		\\
			&			&			\\
8.41$\times$10$^{9}$	&	15		& Measured from the maps of\\
			&			& Hardcastle et al. 1996\\
1.42$\times$10$^{9}$	&	73		& Measured from the maps of\\
			&			& Hardcastle et al. 1996\\ \hline

\end{tabular}
\caption{Counter-jet infrared and radio fluxes.} \label{counter}
\end{center}
\end{table}

\section{Acknowledgments}

We would like to thank Didier Fraix-Burnet for kindly allowing us to
use his I-band images of 3C\,66B. The ISOCAM data presented in this
paper was analysed using `CIA', a joint development by the ESA
Astrophysics Division and the ISOCAM Consortium. The ISOCAM Consortium
is led by the ISOCAM PI, C. Cesarsky, Direction des Sciences de la
Mati\`{e}re, C.E.A., France.

\label{lastpage}

\end{document}